\documentclass[twocolumn]{aastex62}
\newcommand{\kms}{km~s$^{-1}$} 
\newcommand{\mum}{$\,\mu$m}

\newcommand{\lsun}{$L_\odot$}
\newcommand{\msun}{$M_\odot$}
\newcommand{\rsun}{$R_\odot$}
\newcommand{\ngc}{NGC\,6334}
\newcommand{\ngci}{{\ngc}\,I}
\newcommand{\ngcf}{{\ngc}\,F}
\newcommand{\ngcimm}{{\ngci}-MM1}
\newcommand{\ngcimmb}{{\ngci}-MM1B}

\newcommand{\mjb}{mJy~beam$^{-1}$}
\newcommand{\jyb}{Jy~beam$^{-1}$}

\newcommand{\msunperyear}{\msun\,yr$^{-1}$}
\newcommand{\stff}{S255IR-NIRS3}
\newcommand{\gtfe}{G358.93-0.03}
\newcommand{\rprotostar}{$R_{\rm proto}$}
\newcommand{\mdot}{$\dot{M}_{\rm acc}$}
\newcommand{\mstar}{$M_{*}$}
\newcommand{\rstar}{$R_{*}$}
\newcommand{\teff}{$T_{\rm eff}$}
\newcommand{\lbol}{$L_{\rm burst}$}
\newcommand{\lacc}{$L_{\rm acc}$}
\newcommand{\lphoto}{$L_{\rm phot}$}

\newcommand{\lbg}{$L_{\rm preburst}$}
\newcommand{\lburstmmone}{$L_{\rm MM1burst}$}
\newcommand{\lpremmone}{$L_{\rm MM1preburst}$}

\newcommand{\rpre}{$r_{\rm preburst}$}
\newcommand{\rburst}{$r_{\rm burst}$}

\received{March 26, 2021}
\revised{April 9, 2021}
\accepted{April 10, 2021}

\shorttitle{Strong mid-infrared emission from \ngci-MM1 outburst}
\shortauthors{Hunter et al.}

\begin{document}

\title{The Extraordinary Outburst in the Massive Protostellar System NGC\,6334\,I-MM1:\\ Strong Increase in Mid-Infrared Continuum Emission }

\correspondingauthor{T. R. Hunter}
\email{thunter@nrao.edu}

\author[0000-0001-6492-0090]{T. R. Hunter}
\affil{National Radio Astronomy Observatory, 520 Edgemont Rd, Charlottesville, VA 22903, USA}
\affil{Center for Astrophysics $\mid$ Harvard \& Smithsonian, Cambridge, MA 02138, USA}

\author[0000-0002-6558-7653]{C. L. Brogan}
\affil{National Radio Astronomy Observatory, 520 Edgemont Rd, Charlottesville, VA 22903, USA}

\author[0000-0001-7378-4430]{J. M. De Buizer}
\affil{SOFIA-USRA, NASA Ames Research Center, MS 232-12, Moffett Field, CA 94035, USA}

\author[0000-0001-5933-824X]{A. P. M. Towner}
\affil{Department of Astronomy, University of Florida, 211 Bryant Space Science Center, P.O. Box 112055, Gainesville, FL 32611, USA}

\author{C. D. Dowell}
\affil{NASA Jet Propulsion Laboratory, California Institute of Technology, 4800 Oak Grove Drive, Pasadena, CA 91109, USA}

\author[0000-0002-1505-2511]{G. C. MacLeod}
\affil{Hartebeesthoek Radio Astronomy Observatory, PO Box 443, Krugersdorp 1740, South Africa}
\affil{The University of Western Ontario, 1151 Richmond Street, London, ON N6A 3K7, Canada}

\author[0000-0001-6091-163X]{B. Stecklum}
\affil{Th{\"u}ringer Landessternwarte Tautenburg, Sternwarte 5, 07778 Tautenburg, Germany}

\author[0000-0001-6725-1734]{C. J. Cyganowski}
\affil{SUPA, School of Physics and Astronomy, University of St. Andrews, North Haugh, St. Andrews KY16 9SS, UK}

\author{S. J. El-Abd}
\affil{Department of Astronomy, University of Virginia, P.O. Box 3818, Charlottesville, VA 22904, USA}

\author[0000-0003-1254-4817]{B. A. McGuire}
\affil{Department of Chemistry, Massachusetts Institute of Technology, Cambridge MA, 02139, USA}
\affil{National Radio Astronomy Observatory, 520 Edgemont Rd, Charlottesville, VA 22903, USA}
\affil{Center for Astrophysics $\mid$ Harvard \& Smithsonian, Cambridge, MA 02138, USA}



\begin{abstract}

In recent years, dramatic outbursts have been identified toward massive protostars via
infrared and millimeter dust continuum and molecular maser emission.  The longest lived
outburst ($>6$\,yr) persists in \ngcimm, a deeply-embedded object with no near-IR counterpart.  Using
FORCAST and HAWC+ on SOFIA, we have obtained the first mid-IR images of this
field since the outburst began.  Despite being undetected in pre-outburst ground-based 18\mum\/
images, MM1 is now the brightest region at all three wavelengths (25, 37, and 53\mum), exceeding the UCHII
region MM3 (\ngcf). Combining the SOFIA data with ALMA imaging at four wavelengths, we
construct a spectral energy distribution of the combination of MM1 and the nearby hot core MM2.  The best-fit Robitaille
radiative transfer model yields a luminosity of $(4.9\pm0.8)\times10^4$\,\lsun.  
Accounting for an estimated pre-outburst luminosity ratio MM1:MM2 = $2.1\pm0.4$, the luminosity of MM1 has increased by a factor of 16.3$\pm$4.4.
The pre-outburst luminosity implies a protostar of mass 6.7\,\msun, which
can produce the ionizing photon rate required to power the pre-outburst
HCHII region surrounding the likely outbursting protostar MM1B.  
The total energy and duration of the outburst exceed the
\stff\/ outburst by a factor of $\gtrsim3$, suggesting a different scale of
event involving expansion of the protostellar photosphere 
(to $\gtrsim$20\,\rsun), thereby supporting a higher accretion rate 
($\gtrsim$0.0023\,\msunperyear) 
and reducing the ionizing photon rate. 
In the grid of hydrodynamic models of \citet{Meyer21}, the combination of outburst luminosity and magnitude (3) places the \ngcimm\/ event in the region of moderate total accretion ($\sim$0.1-0.3\,\msun) and hence long duration ($\sim$40-130\,yr).

\end{abstract}


\section{Introduction} 

The realization that protostellar emission can be highly variable began with the recognition that the 6-magnitude optical brightening of FU\,Ori arose from a young stellar object (YSO) rather than a nova \citep{Herbig66}.   Similar scale outbursts have since been found in over two dozen Class I and II low-mass protostars \citep[][and references therein]{Gramajo14}, along with the first detection in a Class 0 object \citep{Safron15}.
Such luminosity outbursts, along with smaller ones, offer strong evidence that stars form via continuous slow accretion punctuated by short bursts of rapid, episodic accretion \citep{Vorobyov15}.  The observed timescales of outbursts span a broad range from several months 
to a few hundred years \citep{Hillenbrand15}. Such behavior has been reproduced in hydrodynamic simulations \citep{MacFarlane19}.   A study of Orion suggests that episodic accretion accounts for $\gtrsim$25\% of a star's mass \citep{Fischer19}, making it an important ingredient in star formation.

The recent outburst in near-IR through millimeter continuum from \stff\/ \citep{Caratti17,Liu18}, corresponding to a $\ge$5.5$\times$ increase in luminosity for $\approx$2~years, proved that massive YSOs also exhibit accretion outbursts.
In mid-2015, with the Atacama Large Millimeter/submillimeter Array (ALMA), we discovered an outburst 
from the deeply-embedded massive protocluster \ngci\/ compared to our earlier 2008 Submillimeter Array (SMA) observations \citep{Hunter17}.  
Because \ngci\/ is relatively nearby \citep[1.3~kpc;][]{Reid14,Chibueze14}, we can resolve the protostars,
including the four massive YSOs: MM1..MM4 \citep{Hunter06,Brogan16}.  The two brightest millimeter dust sources, MM1 and MM2, contain extremely line-rich hot molecular cores \citep{McGuire17,Bogelund18,ElAbd19}; MM3 corresponds to the more evolved ultracompact (UC)HII region NGC6334F that is bright at mid-IR/cm wavelengths \citep{DePree95}. MM1, MM2, and MM4 all drive bipolar outflows (Fig.~\ref{introFigure}), while MM1 and MM2 each harbor a hypercompact (HC)HII region \citep[MM1B and MM2B, respectively,][]{Brogan18}.

The \ngci\/ millimeter outburst, centered on MM1B, 
was accompanied by an unprecedented simultaneous flaring 
in multiple maser species monitored by the Hartebeesthoek Radio Astronomy Observatory (HartRAO) 26\,m telescope beginning in January 2015 \citep{MacLeod18}, signaling that an accretion event had begun.   
Accretion outbursts sometimes fade quickly, such as the recent heating event in the protocluster 
\gtfe\/ \citep{Burns20}, which flared in
numerous methanol maser lines never before seen \citep{Breen19,MacLeod19,Brogan19}. 
In contrast, in continued ALMA and HartRAO monitoring of \ngci, the elevated dust continuum and maser emission  have persisted, while Karl G. Jansky Very Large Array (VLA) imaging revealed that strong 6.7\,GHz methanol maser emission had arisen in the vicinity of MM1 as a result of the outburst \citep{Hunter18}.
Here we present the first mid-IR imaging since the event began, along with contemporaneous multi-band ALMA (sub)millimeter measurements, which together provide a more accurate outburst luminosity.

\section{Observations} 

\begin{deluxetable*}{lcccccc}   
\tablewidth{0pc}
\setlength{\tabcolsep}{0.1cm}
\setlength{\tabcolsep}{0.5mm}
\tablecaption{Observing and imaging parameters\label{obstable}}  
\tablehead{ & \multicolumn{2}{c}{SOFIA} & \multicolumn{4}{c}{ALMA} \\ \colhead{Parameter} & \colhead{FORCAST} & 
\colhead{HAWC+} & \colhead{Band 4} & \colhead{Band 7} & \colhead{Band 8} & \colhead{Band 9} }
\startdata
Project code          &  07\_0156\_1    &  GTO 70\_0609\_13         & 2017.1.00661.S & 2017.1.00661.S  & 2017.1.00370.S & 2017.1.00717.S\\
Configuration(s)      & ...   & ...   &  C43-6 \& C43-3  &  C43-5 \& C43-2 &  C43-4  &  C43-3 \\
Observation date(s)   & 2019-07-09, -10  & 2018-07-14                
& 6 executions\tablenotemark{a} & 4 executions\tablenotemark{b}      & 2018-09-12 & 2018-08-28\\
Exposure time (sec)    &  1781, 1611       & 448      &  10215    &  5149   &  2118   & 2365 \\
Flux calibrator & ... & ... & J1617$-$5848 & J1924$-$2914\tablenotemark{b} & J1924$-$2914 & J2253$+$1608 \\
Gain calibrator & ... & ... & J1713$-$3418 & J1717$-$3342 & J1733$-$3722 & J1733$-$3722 \\
Wavelength(s) $\lambda_1, \lambda_2$ ($\mu$m)  & 25.3, 37.1          &  53                       & 2173           & 1005            & 758 & 432 \\
Projected $uv$-range (kilo$\lambda$) &  \nodata     & \nodata                   & 13 -- 1218     & 13 -- 1380      & 16 -- 1613 & 19 -- 1746 \\
Robust weighting & \nodata & \nodata & 0.5 & 0.5 & 0.8 & 0.7 \\
Net continuum BW (MHz) & \nodata & \nodata & 654 & 404 & 105 & 326 \\
rms noise (Jy beam$^{-1}$) $\lambda_1$, $\lambda_2$ & 0.056, 0.094  & 0.34  & 0.00009  & 0.000085  & 0.006 & 0.05 \\
\enddata
\tablenotetext{a}{Observed in configuration C43-6 on 2017-12-03, 2017-12-07, and twice on 2018-01-04, plus twice in C43-3 on 2018-04-23.}
\tablenotetext{b}{Observed in configuration C43-5 on 2018-01-23 and 2018-01-14, and in configuration C43-2 on 2018-05-09 and 2018-05-25; the first execution employed J1517$-$2422 as  flux calibrator.}
\end{deluxetable*}

\subsection{SOFIA}
We used the Stratospheric Observatory for Infrared Astronomy \citep[SOFIA,][]{Temi14} during Cycle~7 to image \ngci\/ with the Faint Object infraRed CAmera for the SOFIA Telescope \citep[FORCAST,][]{Herter2012}. 
Observations were performed in the asymmetric chop-and-nod imaging mode C2NC2. This field was also observed with the High-resolution Airborne Wideband Camera-plus \citep[HAWC+,][]{Harper2018} in Lissajous scanning mode during Guaranteed Time Observations.  Further details are listed in Table~\ref{obstable}.  We used the Level 4 and Level 3 archival data products for FORCAST and HAWC+, respectively.  To facilitate closer comparison with higher resolution data,
we also generated a deconvolved version of the FORCAST 25\mum\/ image with $\approx 2''$ resolution. 

\subsection{ALMA}

ALMA observations with the 12m array were obtained with comparable angular resolution in Bands 4, 7, 8, and 9 (2.2, 1.0, 0.76, and 0.43\,mm) within 8 months during ALMA Cycle 5 (Table~\ref{obstable}).  We calibrated the data with the Cycle 7 release of the ALMA pipeline. \ngci\/ is an extremely line-rich source (\S1).  The method described in \citet{Brogan18} was employed to identify line-free channels; the resulting net continuum bandwidths are listed in Table~\ref{obstable}. Iterative self-calibration was performed on the continuum data to improve the image quality \citep{Brogan18self}. The Briggs robust weighting parameter (smaller values down-weight shorter $uv$-spacings) was used to achieve similar angular-scale sensitivity between the different bands beyond that afforded by the projected uv-ranges of the data (see Table~\ref{obstable}); these ALMA data are not sensitive to {\em smooth} emission larger than about $8\arcsec$. Spectral cubes of SiO (3-2) and CS (6-5) with 2\,\kms\/ channels were created from the Band 4 and 7 continuum-subtracted data, respectively. With the exception of the C43-6 Band 4 data \citep{McGuire18,Xue19,Ligterink20}, results from these ALMA datasets have not previously been published.



\subsection{HartRAO}
\label{hartrao_obs}

We compiled the latest maser monitoring spectra of \ngci\/ from the HartRAO 26\,m telescope. Observation details are provided by \citet{MacLeod18}.  For each maser transition, we extracted a time series of the intensity in the velocity channel closest to the LSR velocity of the thermal gas ($-7.25$\,\kms), which also represents the peak methanol maser emission from MM1 \citep{Hunter18}.

\subsection{Archival data and astrometry}
\label{astrometry}
To establish accurate astrometry for the infrared images, we extracted an archival pre-outburst $0\farcs6$
$K_s$ image from the High Acuity Wide field K-band Imager (HAWKI) on the Very Large Telescope (VLT) on
2012-Jun-09. Applying 2MASS photometric calibration, the faintest detected object is $\sim$21\,mag. Despite the large difference in wavelength, the morphological features
from IRS-I-1 and IRS-I-3 are remarkably similar to the 1998 18\mum\/ image from the Cerro Tololo
Inter-American Observatory (CTIO) 4\,m telescope \citep[Fig.~\ref{introFigure}a;][]{DeBuizer00}.  
We calibrated the HAWKI image astrometry to {\it Gaia}~DR2 \citep{Lindegren2018}, then aligned the CTIO image to it.
We then aligned the three subfields observed in 1999 at W.M.~Keck Observatory \citep{DeBuizer02} to the CTIO image.
Next, we aligned the FORCAST images to the CTIO image, and finally the HAWC+ image to the 37\mum\/ image.
We also extracted the 70\mum\/ PACS image \citep{Tige17} from the {\it Herschel} archive and aligned it to the HAWC+ image.
Table~\ref{fluxtable} lists the shifts applied to each image.  
Finally, we also extracted 3.4 and 4.6\mum\/ images from the (NEO)WISE survey covering pre- and post-outburst \citep{Mainzer2011}, and $K_s$ images from the VVV/VVVX survey \citep{Minniti2010}, all post-outburst.

\section{Results}

%
%
%


\begin{deluxetable*}{lccccc}   
\tablewidth{0pc}
\tablecolumns{6}
\tablecaption{Multi-wavelength astrometry and photometry\label{fluxtable}}  
\tablehead{\colhead{Observatory} & \colhead{Mean wavelength} &  \colhead{Resolution}  &  \colhead{Shift applied}  &  \colhead{Flux density\tablenotemark{a}}  & \colhead{Mean epoch} \\
 & \colhead{(\mum\/)} & \colhead{($\arcsec$)} & \colhead{$\Delta$RA($\arcsec$), $\Delta$Dec($\arcsec$)} & \colhead{(Jy)} & \colhead{}
 }
\startdata
\cutinhead{Data taken prior to the outburst}
VLT   & 2.15  &  0.6  &  0.0, 0.0\tablenotemark{b}  &  $<5.3\times 10^{-6}$   &  2012.44 \\ 
Keck  & 10  & 0.33  &  +1.08, $-$1.59 & 0.13$\pm$0.01 & 1999.32 \\
CTIO  & 18  &  1.0  & +1.65, +0.14    & 3.49$\pm$0.57    &  1998.50 \\
Keck  & 18  &  0.41  & +1.08, $-$1.59   &  3.43$\pm$0.35  & 1999.32 \\ 
{\it Herschel} PACS (HOBYS)   & 70  & 5.9   & +0.0, $-2.0$ & 2260$\pm$452  & 2010.92 \\
SMA\tablenotemark{c,d} (VEX + EXT) & 878  & 0.54$\times$0.28, +9$^\circ$  &  +0.12, +0.12  & 14.6$\pm$1.6  & 2008.12 \\
SMA\tablenotemark{c} (VEX) & 1326 & 0.80$\times$0.34, +18$^\circ$  & +0.06, +0.4  & 4.0$\pm$0.41   & 2008.63 \\
\cutinhead{Data taken since the outburst began in 2015}
VVV/VVVX  & 2.15   &  $0.7-1.0$  & 0.0, 0.0    &  $<4.2\times 10^{-6}$  &  2017.15 \\
SOFIA FORCAST & 25.3 &  3.2  & +0.5, +0.5     & 1014$\pm$129  & 2019.52 \\
SOFIA FORCAST & 37.1 &  3.5  & +0.25, +0.25   & 3604$\pm$447   & 2019.52 \\
SOFIA HAWC+ & 53   &  5.6  & +0.5, +0.5     & 7661$\pm$1246  & 2018.53 \\
ALMA & 432   & 0.31$\times$0.17, $-76^\circ$  &  0.0, 0.0  & 346$\pm$51    & 2018.65 \\
ALMA & 758   & 0.31$\times$0.19, +89$^\circ$  &  0.0, 0.0  & 65.5$\pm$9.7  & 2018.70 \\
ALMA & 1005  & 0.33$\times$0.28, +61$^\circ$  &  0.0, 0.0\tablenotemark{e}  & 31.7$\pm$3.2  & 2018.38 \\
ALMA & 2173  & 0.32$\times$0.24, $-81^\circ$  &  0.0, 0.0  & 2.90$\pm$0.29 & 2018.08 \\
\enddata
\tablenotetext{a}{Value from the apertures shown in Fig.~\ref{midIR}, except the Keck 10 and 18\mum\/ measurements, which are only for IRS-I-2.}
\tablenotetext{b}{Astrometric reference for IR images; the 70\mum\/ astrometry was bootstrapped from point sources in the shifted SOFIA 53\mum\/ image.}
\tablenotetext{c}{Data from \citet{Hunter17}}
\tablenotetext{d}{Augmented with extended configuration data from 2007-Aug-23 (2007A-S007) for a projected uv range = 30-577\,k$\lambda$.}
\tablenotetext{e}{Astrometric reference for SMA images}
\end{deluxetable*}

\subsection{Pre and Post-outburst Continuum Morphology}
\label{forcast}

\begin{figure*} 
\includegraphics[width=1.0\linewidth]{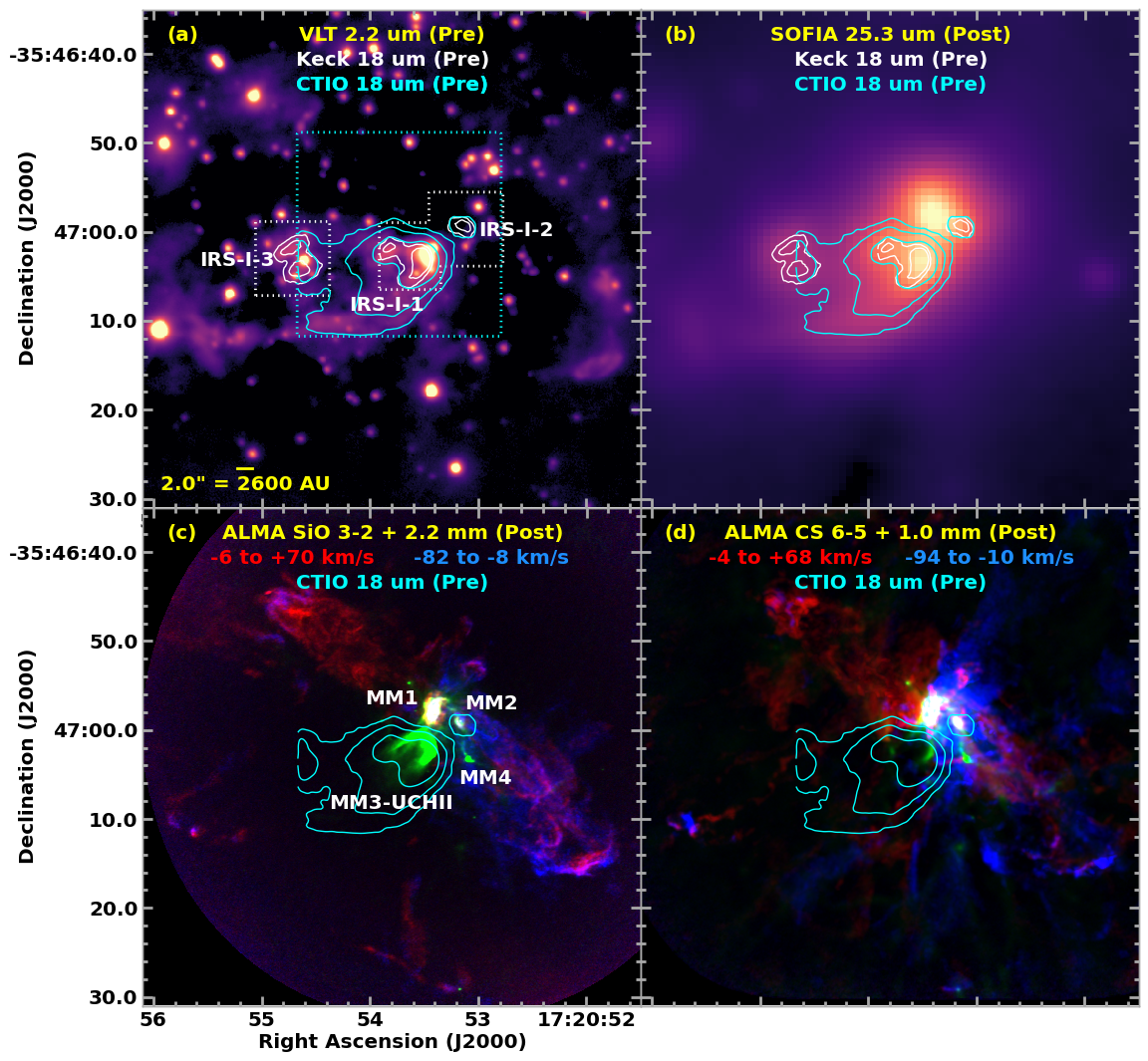}
\centering
\caption{\label{introFigure}\small In each panel, the top label indicates the colorscale image, while (``Pre'') or (``Post'') denotes the time relative to the 2015 outburst. For (c) and (d), RGB is mapped to the integrated redshifted emission of the indicated molecular transitions, continuum, and the integrated blueshifted emission. Each panel is overlaid with cyan CTIO 18\mum\/ contours (smooothed to $1\farcs2$).  (a) and (b) are also overlaid with white Keck 18\mum\/ contours (smoothed to $0\farcs5$); dotted lines in (a) denote their four fields of view. The primary mid-IR and millimeter sources are labeled in (a) and (c), respectively. 
}
\end{figure*}

\begin{figure*}  
\includegraphics[width=1.0\linewidth]{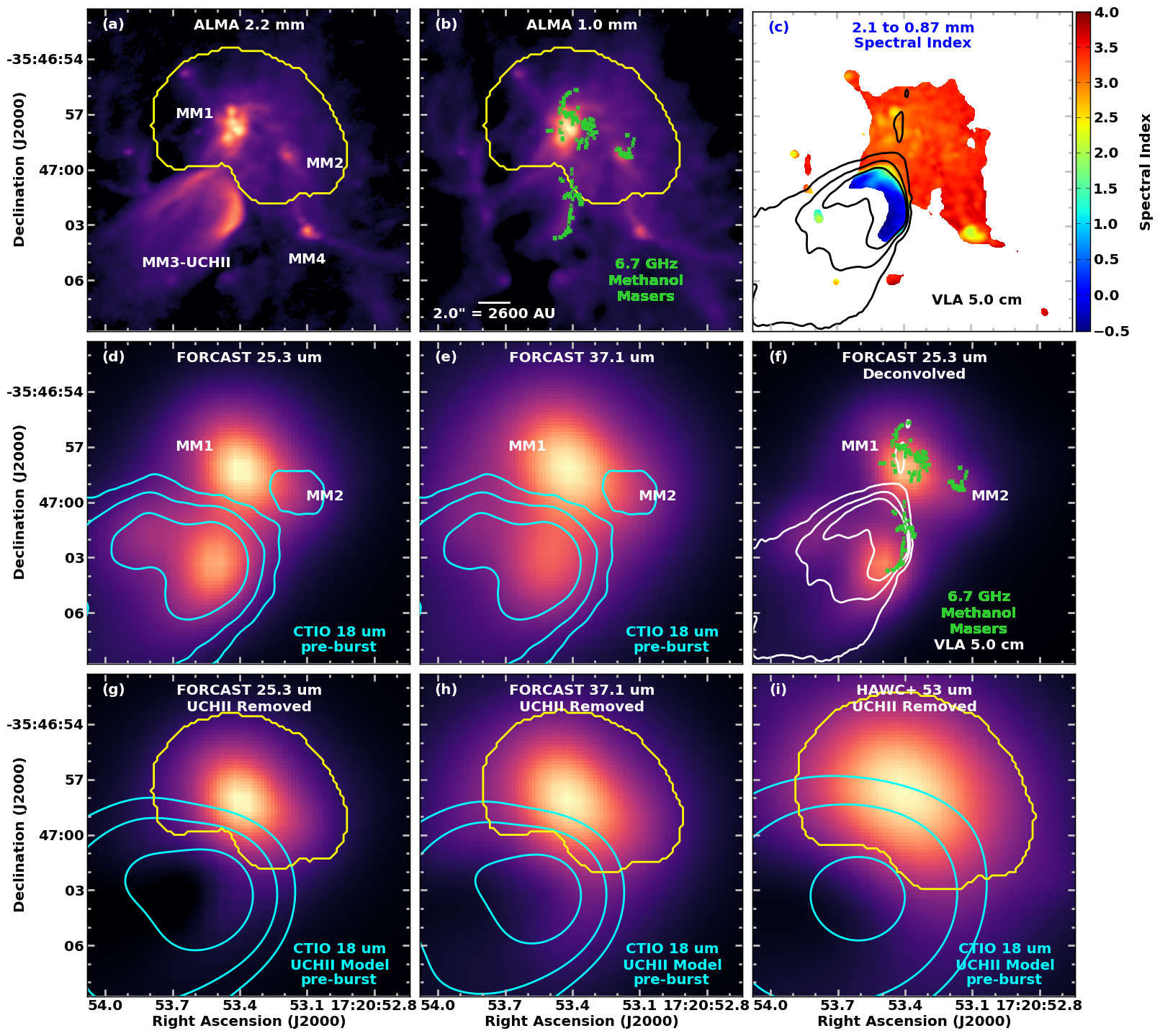}
\centering
\caption{\label{midIR}\small (a) ALMA 2.2\,mm image with the four primary millimeter sources labeled; (b) ALMA 1.0\,mm image; both (a) and (b) show the ALMA photometry aperture as a yellow contour. (c) spectral index image created from the images in (a) and (b), with black VLA 5\,cm contours (0.15, 4.0, 13.2\,\mjb) from \citet{Hunter18}.  (d) and (e) FORCAST 25 and 37\mum\/ images, overlaid with CTIO 18\mum\/ contours (smoothed to $1\farcs2$; 0.7, 1.75, 5.6\,\jyb). (f) deconvolved 25\mum\/ image ($\sim 2\arcsec$ resolution) with the 5~cm contours from (c). Fitted 6.7\,GHz methanol maser positions from \citet{Hunter18} are shown on panels (b) and (f). (g-i) 25, 37, and 53\mum \/ images with the UCHII model removed, cyan contours of the 18\mum\/ image smoothed to the SOFIA image resolutions, and photometry apertures in yellow (scaled in size as described in \S\ref{photometry}). All data were obtained post-outburst except the 18\mum\/ contours in (d,e,g,h,i). Angular resolutions are provided in Table~\ref{fluxtable}.}
\end{figure*}

Prior to the 2015 outburst, mid-IR images of \ngci\/ were dominated by dust emission from
the UCHII region MM3 (IRS-I-1; Fig.~\ref{introFigure}a). 
Aside from MM3, faint emission at 10 and
18\mum\/ was detected toward and slightly southwest of MM2 in the CTIO and Keck images
(IRS-I-2; Fig.~\ref{introFigure}a,b).  The elongation of this feature coincides with
high-velocity blueshifted emission in the ALMA images of SiO 3-2 and CS 6-5  (Fig.~\ref{introFigure}c,d) 
leading toward H$_2$ knot C \citep{Eisloffel00}, 
providing another example of mid-IR emission tracing outflow cavities in massive star-forming regions
\citep{DeBuizer17}.
MM1 is undetected in pre-burst mid-IR imaging \citep{DeBuizer00}, with a 4$\sigma$ upper limit of 0.36\,Jy at 18\mum\/ (see Fig.~\ref{midIR}d contours).

Figures~\ref{midIR}a,b show ALMA continuum images of \ngci, highlighting the four primary protocluster members (MM1-MM4).  The UCHII region MM3 is prominent at 2.2~mm due to its free-free emission, but becomes less apparent at 1.0~mm.  The spectral index ($\alpha$) image ($S_{\nu}\propto \nu^{\alpha}$) formed from these images (Fig.~\ref{midIR}c) demonstrates that MM1, MM2, and MM4 are dominated by dust emission, with the most optically thick sources ($\alpha$ approaching +2) being MM4 and locations within MM1.  In contrast, MM3 shows $\alpha\sim -0.1$ typical of optically thin free-free emission.

As of mid-2019, MM1 is the brightest source at 25 and 37\mum\/, outshining MM3 (Fig.~\ref{midIR}d,e; this is also true in the mid-2018 native 53\mum\/ HAWC+ image, not shown).  This remarkable result confirms a large increase in luminosity, first inferred from millimeter data \citep{Hunter17}. 
Separated by only $3''$, MM1 and MM2 are
difficult to distinguish in the native resolution FORCAST images.  However, the deconvolved 25\mum\/ image (Fig.~\ref{midIR}f) demonstrates that both objects contribute at this wavelength.  Interestingly, the 25\mum\/ emission peaks not at the center of MM1, where the column density is likely highest, but rather $\sim$1$''$ southwest toward a cluster of 6.7\,GHz masers that were not present prior to the outburst \citep{Hunter18}.  The 25\mum\/ emission also extends northward following an arc of masers along the N-S outflow \citep{McGuire18, Brogan18, Chibueze21} and westward, including the MM2 area, supporting the conclusion of \citet{DeBuizer12} that these masers trace mid-IR-bright outflow cavities surrounding massive protostars.  The coincidence of all the methanol masers with 25\mum\/ emission also reflects the expectation of pumping models that require $T_{\rm dust}\gtrsim120$\,K \citep{Cragg05}.

\subsection{Multiwavelength Photometry}
\label{photometry}




The millimeter-wavelength flux ratio of MM1:MM2 increased from \rpre=2.1 in the SMA\footnote{The Submillimeter Array is a joint project between the Smithsonian Astrophysical Observatory and the Academia Sinica Institute of Astronomy and Astrophysics and is funded by the Smithsonian Institution and the Academia Sinica.} 0.87\,mm data to \rburst=10.5 in the ALMA 1.0\,mm data.  However, because MM1 and MM2 cannot be clearly separated in the SOFIA images, the most accurate spectral energy distribution (SED) we can produce is for the combination of the two sources.   Thus, we used the ALMA images to define a polygonal aperture that generally follows a low intensity contour of elliptical shape covering MM1 and MM2 but avoiding the bright northern edge of MM3 (Fig.~\ref{midIR}a,b).
Primary beam correction was applied to all ALMA images prior to measuring flux densities. 
The ALMA flux density uncertainties are the estimated rms noise in the aperture 
(rms per beam $\times\sqrt{\rm beams~in~aperture}$) added in quadrature with the absolute flux uncertainty (10\% for Bands 4 and 7; 15\% for Bands 8 and 9). 

Considering the coarser resolution of the mid-IR images, we needed to model and remove the UCHII region emission to avoid contamination.
To construct a model, we masked the non-UCHII emission in the CTIO 18\mum\/ image, then smoothed it to the resolution of each of the longer wavelength images (Fig.~\ref{midIR}g-i).  Next, each smoothed 18\mum\/ image was iteratively scaled upward to minimize the residuals when subtracting it from the corresponding longer wavelength image. 
For the 25, 37, 53, and 70 \mum\/ images, we found optimal scale factors of 3.33 ($\alpha_{18-25}=3.53$), 7.0 ($\alpha_{18-37}=2.7$), 12 ($\alpha_{18-53}=2.3$), and 2.5, respectively.  These values are comparable to intermediate values of $\alpha_{19-37}$ computed for massive protostar models by \citet{SOMA2020}, and the 37:20\mum\/ and 70:20\mum\/ flux ratios of the cometary HII region M17UC1 \citep[5.7 and 2.25:][]{Lim2020}.

SOFIA and \emph{Herschel} values in Table~\ref{fluxtable} are measured from ``UCHII removed'' images; to account for the larger 37, 53, and 70\mum\/ beams, we expanded the photometric aperture radially from its centroid by the relative increase in beam radius compared to 25\mum\/ ($0\farcs15$, $1\farcs2$, and $1\farcs35$ respectively, see expanded apertures in Fig.~\ref{midIR}h,i).  Removing the UCHII region reduces the measured flux densities by 9.2\%, 4.8\%, 12\%, and 13\% at 25, 37, 53 and 70\mum, respectively.  The photometric uncertainty due to astrometric error was estimated by shifting the aperture by 1 pixel in each of the four cardinal directions, repeating the photometry, then computing the rms of these four possible differences with respect to the nominal flux density: it is 5.1\%, 5.5\%, 1.9\%, and 4.5\% at 25, 37, 53, and 70 \mum, respectively. The total flux density uncertainties -- 12\% at 25 and 37 \mum, 16\% at 53 \mum, and 20\% at 70\mum\/ -- include the estimated noise in the aperture, photometric uncertainty due to astrometric error, UCHII model subtraction uncertainty (half the subtracted contribution), and flux calibration uncertainty \citep[10\% for FORCAST, 15\% for HAWC+, 5\% for PACS,][]{Lim2020,Chuss19,Balog14}, added in quadrature. The total uncertainty at 70\mum\/ includes an additional term for the background subtraction.  

No emission was detected toward MM1 or MM2
in the HAWKI image, with upper limits of 2.65 and 5.3\,$\mu$Jy, respectively. In the (NEO)WISE images,  MM1 is within the saturated PSF core of MM3 in both bands; subtracting pre-outburst from post-outburst images did not reveal any residual toward MM1.  Stacking all the $K_s$ band VVV/VVVX images yields 
an upper limit of 4.2\,$\mu$Jy near both positions.

\subsection{Stability of the Post-outburst Emission}

In order to assess the stability of the continuum emission from \ngci\/ over the span of the post-outburst data, we use the maser light curves from HartRAO.
Among the four species shown in Fig.~\ref{sed}a, the Class II methanol masers (6.7 and 12.2\,GHz) are pumped by mid-IR photons \citep{Cragg05} and may be a good proxy for detecting changes in this radiation. Prior to the January 2015 outburst, the methanol masers arose only from MM2 and MM3 \citep{Green05}. The 12.2\,GHz flare lasted only a few months, but the 6.7\,GHz flare persists.  During the 1.5\,yr interval spanning our ALMA and SOFIA observations (Table~\ref{obstable}), the 6.7\,GHz maser flux density remained constant over 44 epochs with a mean and standard deviation of $716 \pm 45$\,Jy.
Computed using {\tt pyMannKendall} \citep{Hussain2019}, the Mann-Kendall normalized test statistic is 0.23 and the Theil-Sen slope estimator is 0.16, both consistent with no trend.  

\subsection{SEDs and SED fitting}

\begin{figure*}  
\includegraphics[width=1\linewidth]{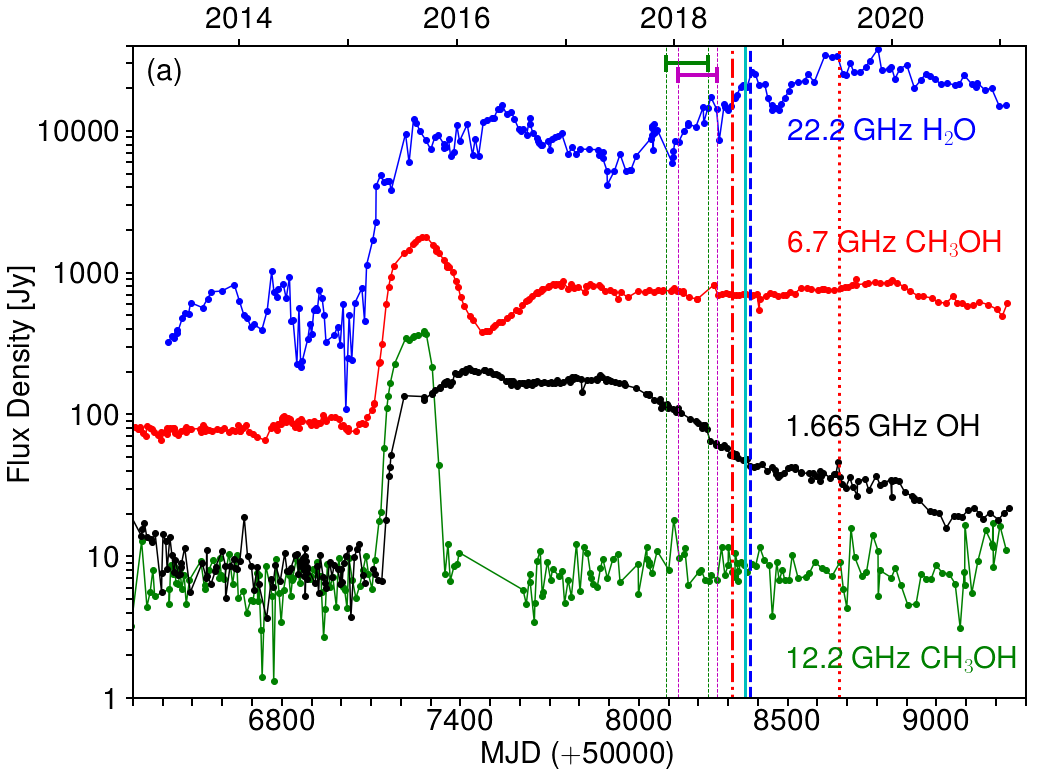}\\
\includegraphics[width=.5\linewidth]{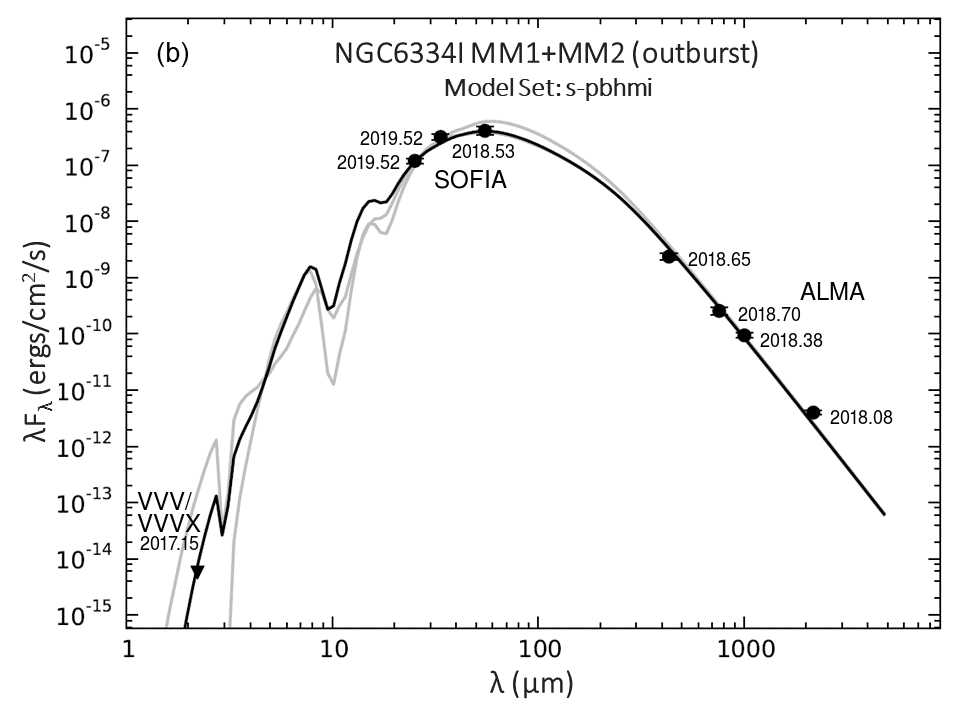}
\includegraphics[width=.5\linewidth]{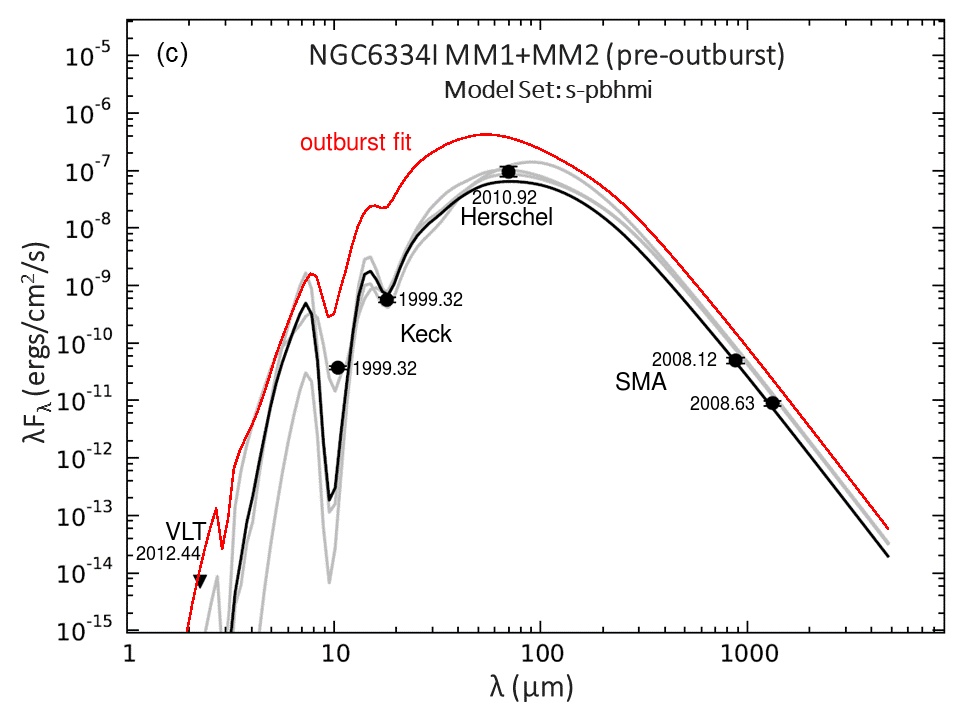}
\begin{center}
\centering
\caption{\label{sed} a) HartRAO light curve of four maser transitions at the LSR velocity ($-7.25$~\kms,  \S~\ref{hartrao_obs}). The dash-dot and dotted vertical lines mark the HAWC+ and FORCAST observation dates, respectively. The dates (or ranges) of the ALMA observations are demarcated: Band 4 (solid green horizontal line with caps signifying first and last dates), Band 7 (solid magenta horizontal line with analogous caps), Band 8 (dashed vertical blue line), and Band 9 (solid vertical cyan line). 
b) The outburst SED (Table~\ref{fluxtable}) overlaid with \citet{Robitaille17} RT models: the best fit is the darkest line.
c) Same as (b) for the pre-outburst SED, with the best fit curve from (b) overlaid in red for comparison.
}
\end{center}
\end{figure*}

Combining observations spanning 18 months into an SED can be problematic for a variable source.
However, massive protostars exhibiting near-IR variability \citep{Kumar16} show similar timescale fluctuations as the 6.7\,GHz masers surrounding deeply-embedded massive protostars \citep{Goedhart14}. 
Also, in a sample of 6.7\,GHz masers with radio continuum counterparts observed on arcsecond scales, the maser luminosity correlates significantly with the inferred ionizing photon rate \citep{Szymczak18b}. Along with theoretical expectations, these observational findings suggest that variations in the radiation field will often manifest in maser variations. Thus, the observed constancy of the methanol maser flux density from \ngcimm\/ during the interval spanning our mid-IR/millimeter observations gives us confidence in constructing the separate SEDs 
in Fig.~\ref{sed}.  

We fit these SEDs to protostellar radiative transfer (RT) models \citep{Robitaille17} using the methodology of \citet{Towner19}. The best fit outburst model (Fig.~\ref{sed}b) includes a power-law envelope with an outflow cavity and a central protostar with bolometric luminosity \lbol\/ = 49000$\pm$8000\,\lsun, where the fractional uncertainty in \lbol\/ has been set to that of the flux density measurement nearest the SED peak (53\mum).
The best fit pre-outburst model (Fig.~\ref{sed}c) comes from the same model family, 
with nearly identical outflow inclination angle but half the opening angle
and has \lbg = 4300$\pm$900\,\lsun (taking the uncertainty from the 70\mum\/ measurement). 
MM1 and MM2 both harbor massive protostars, with hot core molecular spectra of comparable complexity and temperature \citep[$T_{\rm exc}$=150-200\,K,][]{ElAbd19}.
Therefore, we use their pre-outburst millimeter flux density ratio, \rpre\/ (\S\ref{photometry}), to estimate their pre-outburst luminosity ratio.
Since the brightness of MM2 has remained unchanged \citep{Hunter17}, we
compute MM1's pre-outburst luminosity as \lpremmone=2900$\pm$600\,\lsun\/ and
its outburst luminosity as \lburstmmone = $47600\pm7800$\,\lsun\/ (assuming 20\% uncertainty on \rpre).  
Thus, the outburst ratio is 16.3$\pm$4.4, consistent with
a magnitude 3 outburst \citep{Meyer21}.

\section{Discussion}

The total luminosity ($L$) of a massive protostar is a combination of photospheric luminosity (\lphoto) and accretion luminosity (\lacc),
\begin{equation}
L =  4 \pi R_{\rm proto}^{2} \sigma T_{\rm eff}^{4} + G M_* \dot{M}_{\rm acc} / R_{\rm proto},
\end{equation}
with the accretion rate (\mdot) increasing during outbursts while having a persistent ``background'' value in between outbursts \citep[][]{Meyer21}.  If we
assume that half of \lpremmone\/ arises from a ZAMS photosphere with solar metallicity, then the progenitor's properties are: mass \mstar=6.7\,\msun, radius \rstar=2.6\,\rsun, and effective temperature \teff=22000\,K \citep{Tout96}, which lies between spectral type B1.5V and B2V \citep{Pecaut13}.  
Its ionizing photon rate, $Q=2.1\times10^{44}$\,ph\,s$^{-1}$ \citep{Diaz98}, slightly
exceeds that required 
\citep[$1.7\times 10^{44}$ ph\,s$^{-1}$, using Eq.~7 of][]{Carpenter90} to produce the 1.3\,cm flux density of the HCHII region MM1B \citep{Brogan16}.  
This comparison suggests that MM1B contributes the bulk of the luminosity of MM1, dominating over the other point source, MM1D.  
Equal luminosity from accretion would require a background rate of $1.8\times10^{-5}$\,\msunperyear, comparable to the steady accretion rates in the 6\,\msun\/ formation model of \citet{Haemmerle2013}. 

If the increase in MM1's luminosity, $\Delta{L}$ = \lburstmmone-\lpremmone $=44700\pm7900$\,\lsun, 
is entirely attributed to increased accretion, the new accretion rate onto the MM1B protostar would be $6\times10^{-4}$ (\rprotostar/2.6\rsun) \msunperyear. 
However, a sudden increase in accretion rate may cause the protostellar photosphere 
to bloat, resulting in
a temporary excursion on the HR diagram to higher \lphoto, larger \rprotostar, lower \teff, and lower $Q$ 
\citep{Meyer19excursions}.
Interestingly, in our first post-outburst 1.3\,cm VLA observation, we detect 
a substantial drop in the free-free emission from MM1B \citep{Brogan18iau}, 
consistent with lower $Q$ (Brogan et al., in prep.). 
A larger \rprotostar\/ requires a larger \mdot\/ to achieve the same \lacc.  For example, if $Q$ dropped by a factor of 4, and \lburstmmone\/ remained split equally between accretion and protostellar luminosity, then the photosphere would have \teff$\approx$16000\,K and \rprotostar$\approx$20\,\rsun, implying an outburst accretion rate of $2.3\times10^{-3}$ \msunperyear. Theoretical support for such a rate comes from a hydrodynamical study of outburst parameters, where this event's combination of $\Delta{L}$ and duration places it in a region of parameter space containing magnitude 3 outbursts \citep[below center in Fig.~5a of][]{Meyer21}.  The total mass accreted in the closest analogue events is $\sim$0.1-0.3\msun, implying a lifetime for the MM1B outburst of $\sim$40-130\,yr. 

Considering the duration of the ongoing outburst ($\Delta{t}>6$\,yr, based on the maser light curve), we compute the total accretion energy ($E_{\rm acc}$) from $\Delta{L}\Delta{t} > (3.2\pm0.6)\times10^{39}$\,J. 
Because $\Delta{t}$ is much greater than the initial rise time
prior to the first ALMA observation (7 months),
the uncertainty in $E_{\rm acc}$ due to unconstrained details of the luminosity growth profile is of less consequence than for \stff\/ and \gtfe\/, which have $E_{\rm acc} = 1.2\times10^{39}$\,J and $2.9\times10^{38}$\,J, respectively \citep{Stecklum21}.  
Thus, the \ngcimmb\/ outburst now exceeds the other outbursts by $\gtrsim$3$\times$ in both duration and energy. This distinction suggests a different magnitude of event, with a timescale more similar to the FU~Ori phenomenon in low mass protostars. While FU~Ori outbursts are typically interpreted in the context of self-luminous disk instabilities \citep{Hartmann96}, an alternate model involves heating and expansion of the outer layers of the star \citep{Larson80,Herbig03}, which could explain the longer decay times.  Our results emphasize the importance of studying outbursts across the protostellar mass range, which may provide new insights on this pervasive phenomenon.  Continuing to monitor the evolution of \ngcimmb\/ from centimeter through mid-IR wavelengths is critical because the variability level and timescale are among the few observables that can constrain the outburst mechanism.  

\bigskip

\smallskip

\bigskip

\smallskip

\acknowledgments

The National Radio Astronomy Observatory is a facility of the National Science Foundation operated under agreement by the Associated Universities, Inc. This paper uses the following ALMA data: ADS/JAO.ALMA\#2017.1.00370.S, \\ ADS/JAO.ALMA\#2017.1.00661.S and \\ ADS/JAO.ALMA\#2017.1.00717.S. ALMA is a partnership of ESO (representing its member states), NSF (USA) and NINS (Japan), together with NRC (Canada) and NSC and ASIAA (Taiwan) and KASI (Republic of Korea), in cooperation with the Republic of Chile. The Joint ALMA Observatory is operated by ESO, AUI/NRAO and NAOJ.  
{This research made use of NASA's Astrophysics Data System Bibliographic Services.} 
Based in part on observations made with the NASA/DLR Stratospheric Observatory for Infrared Astronomy (SOFIA). SOFIA is jointly operated by the Universities Space Research Association, Inc. (USRA), under NASA contract NNA17BF53C, and the Deutsches SOFIA Institut (DSI) under DLR contract 50 OK 0901 to the University of Stuttgart. Financial support for this work was provided by NASA through award \#07\_0156 issued by USRA. Based in part on observations collected at the European Organisation for Astronomical Research in the Southern Hemisphere under ESO programme 089.C-0852(A).
The Hartebeesthoek telescope is operated by the South African Radio Astronomy Observatory, which is a facility of the National Research Foundation, an agency of the Department of Science and Innovation. 
This work has made use of data from the European Space Agency (ESA) mission
{\it Gaia} (\url{https://www.cosmos.esa.int/gaia}), processed by the {\it Gaia}
Data Processing and Analysis Consortium (DPAC,
\url{https://www.cosmos.esa.int/web/gaia/dpac/consortium}). Funding for the DPAC
has been provided by national institutions, in particular the institutions
participating in the {\it Gaia} Multilateral Agreement.  This research made use of APLpy, an open-source plotting package for Python \citep{aplpy2012}.






\bibliography{bibliography}

\begin{thebibliography}{}
\expandafter\ifx\csname natexlab\endcsname\relax\def\natexlab#1{#1}\fi
\providecommand{\url}[1]{\href{#1}{#1}}

\bibitem[{{Balog} {et~al.}(2014){Balog}, {Muzerolle}, {Flaherty}, {Detre},
  {Bouwmann}, {Furlan}, {Gutermuth}, {Juhasz}, {Bally}, {Nielbock}, {Klaas},
  {Krause}, {Henning}, \& {Marton}}]{Balog14}
{Balog}, Z., {Muzerolle}, J., {Flaherty}, K., {et~al.} 2014, \apjl, 789, L38

\bibitem[{{B{\o}gelund} {et~al.}(2018){B{\o}gelund}, {McGuire}, {Ligterink},
  {Taquet}, {Brogan}, {Hunter}, {Pearson}, {Hogerheijde}, \& {van
  Dishoeck}}]{Bogelund18}
{B{\o}gelund}, E.~G., {McGuire}, B.~A., {Ligterink}, N. F.~W., {et~al.} 2018,
  \aap, 615, A88

\bibitem[{{Breen} {et~al.}(2019){Breen}, {Sobolev}, {Kaczmarek}, {Ellingsen},
  {McCarthy}, \& {Voronkov}}]{Breen19}
{Breen}, S.~L., {Sobolev}, A.~M., {Kaczmarek}, J.~F., {et~al.} 2019, \apjl,
  876, L25

\bibitem[{{Brogan} {et~al.}(2016){Brogan}, {Hunter}, {Cyganowski}, {Chandler},
  {Friesen}, \& {Indebetouw}}]{Brogan16}
{Brogan}, C.~L., {Hunter}, T.~R., {Cyganowski}, C.~J., {et~al.} 2016, \apj,
  832, 187

\bibitem[{{Brogan} {et~al.}(2018{\natexlab{a}}){Brogan}, {Hunter}, \&
  {Fomalont}}]{Brogan18self}
{Brogan}, C.~L., {Hunter}, T.~R., \& {Fomalont}, E.~B. 2018{\natexlab{a}},
  ArXiv e-prints, arXiv:1805.05266

\bibitem[{{Brogan} {et~al.}(2018{\natexlab{b}}){Brogan}, {Hunter}, {MacLeod},
  {Chibueze}, \& {Cyganowski}}]{Brogan18iau}
{Brogan}, C.~L., {Hunter}, T.~R., {MacLeod}, G., {Chibueze}, J.~O., \&
  {Cyganowski}, C.~J. 2018{\natexlab{b}}, in IAU Symposium, Vol. 336,
  Astrophysical Masers: Unlocking the Mysteries of the Universe, ed.
  A.~{Tarchi}, M.~J. {Reid}, \& P.~{Castangia}, 255--258

\bibitem[{{Brogan} {et~al.}(2018{\natexlab{c}}){Brogan}, {Hunter},
  {Cyganowski}, {Chibueze}, {Friesen}, {Hirota}, {MacLeod}, {McGuire}, \&
  {Sobolev}}]{Brogan18}
{Brogan}, C.~L., {Hunter}, T.~R., {Cyganowski}, C.~J., {et~al.}
  2018{\natexlab{c}}, \apj, 866, 87

\bibitem[{{Brogan} {et~al.}(2019){Brogan}, {Hunter}, {Towner}, {McGuire},
  {MacLeod}, {Gurwell}, {Cyganowski}, {Brand}, {Burns}, {Caratti o Garatti},
  {Chen}, {Chibueze}, {Hirano}, {Hirota}, {Kim}, {Kramer}, {Linz}, {Menten},
  {Remijan}, {Sanna}, {Sobolev}, {Sridharan}, {Stecklum}, {Sugiyama}, {Surcis},
  {Van der Walt}, {Volvach}, \& {Volvach}}]{Brogan19}
{Brogan}, C.~L., {Hunter}, T.~R., {Towner}, A.~P.~M., {et~al.} 2019, \apjl,
  881, L39

\bibitem[{{Burns} {et~al.}(2020){Burns}, {Sugiyama}, {Hirota}, {Kim},
  {Sobolev}, {Stecklum}, {MacLeod}, {Yonekura}, {Olech}, {Orosz}, {Ellingsen},
  {Hyland}, {Caratti o Garatti}, {Brogan}, {Hunter}, {Phillips}, {van den
  Heever}, {Eisl{\"o}ffel}, {Linz}, {Surcis}, {Chibueze}, {Baan}, \&
  {Kramer}}]{Burns20}
{Burns}, R.~A., {Sugiyama}, K., {Hirota}, T., {et~al.} 2020, NatAs, 4, 506

\bibitem[{{Caratti o Garatti} {et~al.}(2017){Caratti o Garatti}, {Stecklum},
  {Garcia Lopez}, {Eisl{\"o}ffel}, {Ray}, {Sanna}, {Cesaroni}, {Walmsley},
  {Oudmaijer}, {de Wit}, {Moscadelli}, {Greiner}, {Krabbe}, {Fischer}, {Klein},
  \& {Iba{\~n}ez}}]{Caratti17}
{Caratti o Garatti}, A., {Stecklum}, B., {Garcia Lopez}, R., {et~al.} 2017,
  NatPh, 13, 276

\bibitem[{{Carpenter} {et~al.}(1990){Carpenter}, {Snell}, \&
  {Schloerb}}]{Carpenter90}
{Carpenter}, J.~M., {Snell}, R.~L., \& {Schloerb}, F.~P. 1990, \apj, 362, 147

\bibitem[{{Chibueze} {et~al.}(2021){Chibueze}, {MacLeod}, {Vorster}, {Hirota},
  {Brogan}, {Hunter}, \& {van Rooyen}}]{Chibueze21}
{Chibueze}, J.~O., {MacLeod}, G.~C., {Vorster}, J.~M., {et~al.} 2021, \apj,
  908, 175

\bibitem[{{Chibueze} {et~al.}(2014){Chibueze}, {Omodaka}, {Handa}, {Imai},
  {Kurayama}, {Nagayama}, {Sunada}, {Nakano}, {Hirota}, \&
  {Honma}}]{Chibueze14}
{Chibueze}, J.~O., {Omodaka}, T., {Handa}, T., {et~al.} 2014, \apj, 784, 114

\bibitem[{{Chuss} {et~al.}(2019){Chuss}, {Andersson}, {Bally}, {Dotson},
  {Dowell}, {Guerra}, {Harper}, {Houde}, {Jones}, {Lazarian}, {Lopez
  Rodriguez}, {Michail}, {Morris}, {Novak}, {Siah}, {Staguhn}, {Vaillancourt},
  {Volpert}, {Werner}, {Wollack}, {Benford}, {Berthoud}, {Cox}, {Crutcher},
  {Dale}, {Fissel}, {Goldsmith}, {Hamilton}, {Hanany}, {Henning}, {Looney},
  {Moseley}, {Santos}, {Stephens}, {Tassis}, {Trinh}, {Van Camp},
  {Ward-Thompson}, \& {HAWC + Science Team}}]{Chuss19}
{Chuss}, D.~T., {Andersson}, B.~G., {Bally}, J., {et~al.} 2019, \apj, 872, 187

\bibitem[{{Cragg} {et~al.}(2005){Cragg}, {Sobolev}, \& {Godfrey}}]{Cragg05}
{Cragg}, D.~M., {Sobolev}, A.~M., \& {Godfrey}, P.~D. 2005, \mnras, 360, 533

\bibitem[{{De Buizer} {et~al.}(2012){De Buizer}, {Bartkiewicz}, \&
  {Szymczak}}]{DeBuizer12}
{De Buizer}, J.~M., {Bartkiewicz}, A., \& {Szymczak}, M. 2012, \apj, 754, 149

\bibitem[{{De Buizer} {et~al.}(2000){De Buizer}, {Pi{\~n}a}, \&
  {Telesco}}]{DeBuizer00}
{De Buizer}, J.~M., {Pi{\~n}a}, R.~K., \& {Telesco}, C.~M. 2000, \apjs, 130,
  437

\bibitem[{{De Buizer} {et~al.}(2002){De Buizer}, {Radomski}, {Pi{\~n}a}, \&
  {Telesco}}]{DeBuizer02}
{De Buizer}, J.~M., {Radomski}, J.~T., {Pi{\~n}a}, R.~K., \& {Telesco}, C.~M.
  2002, \apj, 580, 305

\bibitem[{{De Buizer} {et~al.}(2017){De Buizer}, {Liu}, {Tan}, {Zhang},
  {Beltr{\'a}n}, {Shuping}, {Staff}, {Tanaka}, \& {Whitney}}]{DeBuizer17}
{De Buizer}, J.~M., {Liu}, M., {Tan}, J.~C., {et~al.} 2017, \apj, 843, 33

\bibitem[{{de Pree} {et~al.}(1995){de Pree}, {Rodriguez}, {Dickel}, \&
  {Goss}}]{DePree95}
{de Pree}, C.~G., {Rodriguez}, L.~F., {Dickel}, H.~R., \& {Goss}, W.~M. 1995,
  \apj, 447, 220

\bibitem[{{Diaz-Miller} {et~al.}(1998){Diaz-Miller}, {Franco}, \&
  {Shore}}]{Diaz98}
{Diaz-Miller}, R.~I., {Franco}, J., \& {Shore}, S.~N. 1998, \apj, 501, 192

\bibitem[{{Eisl{\"o}ffel} {et~al.}(2000){Eisl{\"o}ffel}, {Smith}, \&
  {Davis}}]{Eisloffel00}
{Eisl{\"o}ffel}, J., {Smith}, M.~D., \& {Davis}, C.~J. 2000, \aap, 359, 1147

\bibitem[{{El-Abd} {et~al.}(2019){El-Abd}, {Brogan}, {Hunter}, {Willis},
  {Garrod}, \& {McGuire}}]{ElAbd19}
{El-Abd}, S.~J., {Brogan}, C.~L., {Hunter}, T.~R., {et~al.} 2019, \apj, 883,
  129

\bibitem[{{Fischer} {et~al.}(2019){Fischer}, {Safron}, \&
  {Megeath}}]{Fischer19}
{Fischer}, W.~J., {Safron}, E., \& {Megeath}, S.~T. 2019, \apj, 872, 183

\bibitem[{{Goedhart} {et~al.}(2014){Goedhart}, {Maswanganye}, {Gaylard}, \&
  {van der Walt}}]{Goedhart14}
{Goedhart}, S., {Maswanganye}, J.~P., {Gaylard}, M.~J., \& {van der Walt},
  D.~J. 2014, \mnras, 437, 1808

\bibitem[{{Gramajo} {et~al.}(2014){Gramajo}, {Rod{\'o}n}, \&
  {G{\'o}mez}}]{Gramajo14}
{Gramajo}, L.~V., {Rod{\'o}n}, J.~A., \& {G{\'o}mez}, M. 2014, \aj, 147, 140

\bibitem[{{Green} {et~al.}(2015){Green}, {Caswell}, \&
  {McClure-Griffiths}}]{Green05}
{Green}, J.~A., {Caswell}, J.~L., \& {McClure-Griffiths}, N.~M. 2015, \mnras,
  451, 74

\bibitem[{{Haemmerl{\'e}} {et~al.}(2013){Haemmerl{\'e}}, {Eggenberger},
  {Meynet}, {Maeder}, \& {Charbonnel}}]{Haemmerle2013}
{Haemmerl{\'e}}, L., {Eggenberger}, P., {Meynet}, G., {Maeder}, A., \&
  {Charbonnel}, C. 2013, \aap, 557, A112

\bibitem[{{Harper} {et~al.}(2018){Harper}, {Runyan}, {Dowell}, {Wirth},
  {Amato}, {Ames}, {Amiri}, {Banks}, {Bartels}, {Benford}, {Berthoud},
  {Buchanan}, {Casey}, {Chapman}, {Chuss}, {Cook}, {Derro}, {Dotson}, {Evans},
  {Fixsen}, {Gatley}, {Guerra}, {Halpern}, {Hamilton}, {Hamlin}, {Hansen},
  {Heimsath}, {Hermida}, {Hilton}, {Hirsch}, {Hollister}, {Hostetter}, {Irwin},
  {Jhabvala}, {Jhabvala}, {Kastner}, {Kov{\'a}cs}, {Lin}, {Loewenstein},
  {Looney}, {Lopez-Rodriguez}, {Maher}, {Michail}, {Miller}, {Moseley},
  {Novak}, {Pernic}, {Rennick}, {Rhody}, {Sandberg}, {Sand ford}, {Santos},
  {Shafer}, {Sharp}, {Shirron}, {Siah}, {Silverberg}, {Sparr}, {Spotz},
  {Staguhn}, {Toorian}, {Towey}, {Tuttle}, {Vaillancourt}, {Voellmer},
  {Volpert}, {Wang}, \& {Wollack}}]{Harper2018}
{Harper}, D.~A., {Runyan}, M.~C., {Dowell}, C.~D., {et~al.} 2018, JAI, 7,
  1840008

\bibitem[{{Hartmann} \& {Kenyon}(1996)}]{Hartmann96}
{Hartmann}, L., \& {Kenyon}, S.~J. 1996, \araa, 34, 207

\bibitem[{{Herbig}(1966)}]{Herbig66}
{Herbig}, G. 1966, VA, 8, 109

\bibitem[{{Herbig} {et~al.}(2003){Herbig}, {Petrov}, \& {Duemmler}}]{Herbig03}
{Herbig}, G.~H., {Petrov}, P.~P., \& {Duemmler}, R. 2003, \apj, 595, 384

\bibitem[{{Herter} {et~al.}(2012){Herter}, {Adams}, {De Buizer}, {Gull},
  {Schoenwald}, {Henderson}, {Keller}, {Nikola}, {Stacey}, \&
  {Vacca}}]{Herter2012}
{Herter}, T.~L., {Adams}, J.~D., {De Buizer}, J.~M., {et~al.} 2012, \apjl, 749,
  L18

\bibitem[{{Hillenbrand} \& {Findeisen}(2015)}]{Hillenbrand15}
{Hillenbrand}, L.~A., \& {Findeisen}, K.~P. 2015, \apj, 808, 68

\bibitem[{{Hunter} {et~al.}(2006){Hunter}, {Brogan}, {Megeath}, {Menten},
  {Beuther}, \& {Thorwirth}}]{Hunter06}
{Hunter}, T.~R., {Brogan}, C.~L., {Megeath}, S.~T., {et~al.} 2006, \apj, 649,
  888

\bibitem[{{Hunter} {et~al.}(2017){Hunter}, {Brogan}, {MacLeod}, {Cyganowski},
  {Chandler}, {Chibueze}, {Friesen}, {Indebetouw}, {Thesner}, \&
  {Young}}]{Hunter17}
{Hunter}, T.~R., {Brogan}, C.~L., {MacLeod}, G., {et~al.} 2017, \apjl, 837, L29

\bibitem[{{Hunter} {et~al.}(2018){Hunter}, {Brogan}, {MacLeod}, {Cyganowski},
  {Chibueze}, {Friesen}, {Hirota}, {Smits}, {Chandler}, \&
  {Indebetouw}}]{Hunter18}
{Hunter}, T.~R., {Brogan}, C.~L., {MacLeod}, G.~C., {et~al.} 2018, \apj, 854,
  170

\bibitem[{{Hussain} \& {Mahmud}(2019)}]{Hussain2019}
{Hussain}, M., \& {Mahmud}, I. 2019, JOSS, 4, 1556

\bibitem[{{Kumar} {et~al.}(2016){Kumar}, {Contreras Pe{\~n}a}, {Lucas}, \&
  {Thompson}}]{Kumar16}
{Kumar}, M.~S.~N., {Contreras Pe{\~n}a}, C., {Lucas}, P.~W., \& {Thompson},
  M.~A. 2016, \apj, 833, 24

\bibitem[{{Larson}(1980)}]{Larson80}
{Larson}, R.~B. 1980, \mnras, 190, 321

\bibitem[{{Ligterink} {et~al.}(2020){Ligterink}, {El-Abd}, {Brogan}, {Hunter},
  {Remijan}, {Garrod}, \& {McGuire}}]{Ligterink20}
{Ligterink}, N. F.~W., {El-Abd}, S.~J., {Brogan}, C.~L., {et~al.} 2020, \apj,
  901, 37

\bibitem[{{Lim} {et~al.}(2020){Lim}, {De Buizer}, \& {Radomski}}]{Lim2020}
{Lim}, W., {De Buizer}, J.~M., \& {Radomski}, J.~T. 2020, \apj, 888, 98

\bibitem[{{Lindegren} {et~al.}(2018){Lindegren}, {Hern{\'a}ndez}, {Bombrun},
  {Klioner}, {Bastian}, {Ramos-Lerate}, {de Torres}, {Steidelm{\"u}ller},
  {Stephenson}, {Hobbs}, {Lammers}, {Biermann}, {Geyer}, {Hilger}, {Michalik},
  {Stampa}, {McMillan}, {Casta{\~n}eda}, {Clotet}, {Comoretto}, {Davidson},
  {Fabricius}, {Gracia}, {Hambly}, {Hutton}, {Mora}, {Portell}, {van Leeuwen},
  {Abbas}, {Abreu}, {Altmann}, {Andrei}, {Anglada}, {Balaguer-N{\'u}{\~n}ez},
  {Barache}, {Becciani}, {Bertone}, {Bianchi}, {Bouquillon}, {Bourda},
  {Br{\"u}semeister}, {Bucciarelli}, {Busonero}, {Buzzi}, {Cancelliere},
  {Carlucci}, {Charlot}, {Cheek}, {Crosta}, {Crowley}, {de Bruijne}, {de
  Felice}, {Drimmel}, {Esquej}, {Fienga}, {Fraile}, {Gai}, {Garralda},
  {Gonz{\'a}lez-Vidal}, {Guerra}, {Hauser}, {Hofmann}, {Holl}, {Jordan},
  {Lattanzi}, {Lenhardt}, {Liao}, {Licata}, {Lister}, {L{\"o}ffler},
  {Marchant}, {Martin-Fleitas}, {Messineo}, {Mignard}, {Morbidelli}, {Poggio},
  {Riva}, {Rowell}, {Salguero}, {Sarasso}, {Sciacca}, {Siddiqui}, {Smart},
  {Spagna}, {Steele}, {Taris}, {Torra}, {van Elteren}, {van Reeven}, \&
  {Vecchiato}}]{Lindegren2018}
{Lindegren}, L., {Hern{\'a}ndez}, J., {Bombrun}, A., {et~al.} 2018, \aap, 616,
  A2

\bibitem[{{Liu} {et~al.}(2020){Liu}, {Tan}, {De Buizer}, {Zhang}, {Moser},
  {Beltr{\'a}n}, {Staff}, {Tanaka}, {Whitney}, {Rosero}, {Yang}, \&
  {Fedriani}}]{SOMA2020}
{Liu}, M., {Tan}, J.~C., {De Buizer}, J.~M., {et~al.} 2020, \apj, 904, 75

\bibitem[{{Liu} {et~al.}(2018){Liu}, {Su}, {Zinchenko}, {Wang}, \&
  {Wang}}]{Liu18}
{Liu}, S.-Y., {Su}, Y.-N., {Zinchenko}, I., {Wang}, K.-S., \& {Wang}, Y. 2018,
  \apjl, 863, L12

\bibitem[{{MacFarlane} {et~al.}(2019){MacFarlane}, {Stamatellos}, {Johnstone},
  {Herczeg}, {Baek}, {Chen}, {Kang}, \& {Lee}}]{MacFarlane19}
{MacFarlane}, B., {Stamatellos}, D., {Johnstone}, D., {et~al.} 2019, \mnras,
  487, 5106

\bibitem[{{MacLeod} {et~al.}(2018){MacLeod}, {Smits}, {Goedhart}, {Hunter},
  {Brogan}, {Chibueze}, {van den Heever}, {Thesner}, {Banda}, \&
  {Paulsen}}]{MacLeod18}
{MacLeod}, G.~C., {Smits}, D.~P., {Goedhart}, S., {et~al.} 2018, \mnras, 478,
  1077

\bibitem[{{MacLeod} {et~al.}(2019){MacLeod}, {Sugiyama}, {Hunter}, {Quick},
  {Baan}, {Breen}, {Brogan}, {Burns}, {Caratti o Garatti}, {Chen}, {Chibueze},
  {Houde}, {Kaczmarek}, {Linz}, {Rajabi}, {Saito}, {Schmidl}, {Sobolev},
  {Stecklum}, {van den Heever}, \& {Yonekura}}]{MacLeod19}
{MacLeod}, G.~C., {Sugiyama}, K., {Hunter}, T.~R., {et~al.} 2019, \mnras, 489,
  3981

\bibitem[{{Mainzer} {et~al.}(2011){Mainzer}, {Bauer}, {Grav}, {Masiero},
  {Cutri}, {Dailey}, {Eisenhardt}, {McMillan}, {Wright}, {Walker}, {Jedicke},
  {Spahr}, {Tholen}, {Alles}, {Beck}, {Brandenburg}, {Conrow}, {Evans},
  {Fowler}, {Jarrett}, {Marsh}, {Masci}, {McCallon}, {Wheelock}, {Wittman},
  {Wyatt}, {DeBaun}, {Elliott}, {Elsbury}, {Gautier}, {Gomillion}, {Leisawitz},
  {Maleszewski}, {Micheli}, \& {Wilkins}}]{Mainzer2011}
{Mainzer}, A., {Bauer}, J., {Grav}, T., {et~al.} 2011, \apj, 731, 53

\bibitem[{{McGuire} {et~al.}(2017){McGuire}, {Shingledecker}, {Willis},
  {Burkhardt}, {El-Abd}, {Motiyenko}, {Brogan}, {Hunter}, {Margul{\`e}s},
  {Guillemin}, {Garrod}, {Herbst}, \& {Remijan}}]{McGuire17}
{McGuire}, B.~A., {Shingledecker}, C.~N., {Willis}, E.~R., {et~al.} 2017,
  \apjl, 851, L46

\bibitem[{{McGuire} {et~al.}(2018){McGuire}, {Brogan}, {Hunter}, {Remijan},
  {Blake}, {Burkhardt}, {Carroll}, {van Dishoeck}, {Garrod}, {Linnartz},
  {Shingledecker}, \& {Willis}}]{McGuire18}
{McGuire}, B.~A., {Brogan}, C.~L., {Hunter}, T.~R., {et~al.} 2018, \apj, 863,
  L35

\bibitem[{{Meyer} {et~al.}(2019){Meyer}, {Haemmerl{\'e}}, \&
  {Vorobyov}}]{Meyer19excursions}
{Meyer}, D.~M.-A., {Haemmerl{\'e}}, L., \& {Vorobyov}, E.~I. 2019, \mnras, 484,
  2482

\bibitem[{{Meyer} {et~al.}(2021){Meyer}, {Vorobyov}, {Elbakyan},
  {Eisl{\"o}ffel}, {Sobolev}, \& {St{\"o}hr}}]{Meyer21}
{Meyer}, D.~M.~A., {Vorobyov}, E.~I., {Elbakyan}, V.~G., {et~al.} 2021, \mnras,
  500, 4448

\bibitem[{{Minniti} {et~al.}(2010){Minniti}, {Lucas}, {Emerson}, {Saito},
  {Hempel}, {Pietrukowicz}, {Ahumada}, {Alonso}, {Alonso-Garcia}, {Arias},
  {Bandyopadhyay}, {Barb{\'a}}, {Barbuy}, {Bedin}, {Bica}, {Borissova},
  {Bronfman}, {Carraro}, {Catelan}, {Clari{\'a}}, {Cross}, {de Grijs},
  {D{\'e}k{\'a}ny}, {Drew}, {Fari{\~n}a}, {Feinstein}, {Fern{\'a}ndez
  Laj{\'u}s}, {Gamen}, {Geisler}, {Gieren}, {Goldman}, {Gonzalez}, {Gunthardt},
  {Gurovich}, {Hambly}, {Irwin}, {Ivanov}, {Jord{\'a}n}, {Kerins}, {Kinemuchi},
  {Kurtev}, {L{\'o}pez-Corredoira}, {Maccarone}, {Masetti}, {Merlo},
  {Messineo}, {Mirabel}, {Monaco}, {Morelli}, {Padilla}, {Palma}, {Parisi},
  {Pignata}, {Rejkuba}, {Roman-Lopes}, {Sale}, {Schreiber}, {Schr{\"o}der},
  {Smith}, {}, {Soto}, {Tamura}, {Tappert}, {Thompson}, {Toledo}, {Zoccali}, \&
  {Pietrzynski}}]{Minniti2010}
{Minniti}, D., {Lucas}, P.~W., {Emerson}, J.~P., {et~al.} 2010, \na, 15, 433

\bibitem[{{Pecaut} \& {Mamajek}(2013)}]{Pecaut13}
{Pecaut}, M.~J., \& {Mamajek}, E.~E. 2013, \apjs, 208, 9

\bibitem[{{Reid} {et~al.}(2014){Reid}, {Menten}, {Brunthaler}, {Zheng}, {Dame},
  {Xu}, {Wu}, {Zhang}, {Sanna}, {Sato}, {Hachisuka}, {Choi}, {Immer},
  {Moscadelli}, {Rygl}, \& {Bartkiewicz}}]{Reid14}
{Reid}, M.~J., {Menten}, K.~M., {Brunthaler}, A., {et~al.} 2014, \apj, 783, 130

\bibitem[{{Robitaille} \& {Bressert}(2012)}]{aplpy2012}
{Robitaille}, T., \& {Bressert}, E. 2012, {APLpy: Astronomical Plotting Library
  in Python}, , , ascl:1208.017

\bibitem[{{Robitaille}(2017)}]{Robitaille17}
{Robitaille}, T.~P. 2017, \aap, 600, A11

\bibitem[{{Safron} {et~al.}(2015){Safron}, {Fischer}, {Megeath}, {Furlan},
  {Stutz}, {Stanke}, {Billot}, {Rebull}, {Tobin}, {Ali}, {Allen}, {Booker},
  {Watson}, \& {Wilson}}]{Safron15}
{Safron}, E.~J., {Fischer}, W.~J., {Megeath}, S.~T., {et~al.} 2015, \apjl, 800,
  L5

\bibitem[{{Stecklum} {et~al.}(2021){Stecklum}, {Wolf}, {Linz}, {Caratti o
  Garatti}, {Schmidl}, {Klose}, {Eisl{\"o}ffel}, {Fischer}, {Brogan}, {Burns},
  {Bayandina}, {Cyganowski}, {Gurwell}, {Hunter}, {Hirano}, {Kim}, {MacLeod},
  {Menten}, {Olech}, {Orosz}, {Sobolev}, {Sridharan}, {Surcis}, {Sugiyama},
  {van der Walt}, {Volvach}, \& {Yonekura}}]{Stecklum21}
{Stecklum}, B., {Wolf}, V., {Linz}, H., {et~al.} 2021, \aap, 646, A161

\bibitem[{{Szymczak} {et~al.}(2018){Szymczak}, {Olech}, {Sarniak}, {Wolak}, \&
  {Bartkiewicz}}]{Szymczak18b}
{Szymczak}, M., {Olech}, M., {Sarniak}, R., {Wolak}, P., \& {Bartkiewicz}, A.
  2018, \mnras, 474, 219

\bibitem[{{Temi} {et~al.}(2014){Temi}, {Marcum}, {Young}, {Adams}, {Adams},
  {Andersson}, {Becklin}, {Boogert}, {Brewster}, {Burgh}, {Cobleigh}, {Culp},
  {De Buizer}, {Dunham}, {Engfer}, {Ediss}, {Fujieh}, {Grashuis}, {Gross},
  {Harmon}, {Helton}, {Hoffman}, {Homan}, {H{\"u}twohl}, {Jakob}, {Jensen},
  {Kaminski}, {Kozarsky}, {Krabbe}, {Klein}, {Lammen}, {Lampater}, {Latter},
  {Le}, {McKown}, {Melchiorri}, {Meyer}, {Miles}, {Miller}, {Miller}, {Moore},
  {Nickison}, {Opshaug}, {Pf{\"u}eller}, {Radomski}, {Rasmussen}, {Reach},
  {Reinacher}, {Roellig}, {Sandell}, {Sankrit}, {Savage}, {Shenoy},
  {Schonfeld}, {Shuping}, {Smith}, {Talebi}, {Teufel}, {Tseng}, {Vacca},
  {Vaillancourt}, {Van Cleve}, {Wiedemann}, {Wolf}, {Zavala}, {Zeile}, {Zell},
  \& {Zinnecker}}]{Temi14}
{Temi}, P., {Marcum}, P.~M., {Young}, E., {et~al.} 2014, \apjs, 212, 24

\bibitem[{{Tig{\'e}} {et~al.}(2017){Tig{\'e}}, {Motte}, {Russeil}, {Zavagno},
  {Hennemann}, {Schneider}, {Hill}, {Nguyen Luong}, {Di Francesco}, {Bontemps},
  {Louvet}, {Didelon}, {K{\"o}nyves}, {Andr{\'e}}, {Leuleu}, {Bardagi},
  {Anderson}, {Arzoumanian}, {Benedettini}, {Bernard}, {Elia}, {Figueira},
  {Kirk}, {Martin}, {Minier}, {Molinari}, {Nony}, {Persi}, {Pezzuto},
  {Polychroni}, {Rayner}, {Rivera-Ingraham}, {Roussel}, {Rygl}, {Spinoglio}, \&
  {White}}]{Tige17}
{Tig{\'e}}, J., {Motte}, F., {Russeil}, D., {et~al.} 2017, \aap, 602, A77

\bibitem[{{Tout} {et~al.}(1996){Tout}, {Pols}, {Eggleton}, \& {Han}}]{Tout96}
{Tout}, C.~A., {Pols}, O.~R., {Eggleton}, P.~P., \& {Han}, Z. 1996, \mnras,
  281, 257

\bibitem[{{Towner} {et~al.}(2019){Towner}, {Brogan}, {Hunter}, {Cyganowski}, \&
  {Friesen}}]{Towner19}
{Towner}, A.~P.~M., {Brogan}, C.~L., {Hunter}, T.~R., {Cyganowski}, C.~J., \&
  {Friesen}, R.~K. 2019, \apj, 875, 135

\bibitem[{{Vorobyov} \& {Basu}(2015)}]{Vorobyov15}
{Vorobyov}, E.~I., \& {Basu}, S. 2015, \apj, 805, 115

\bibitem[{{Xue} {et~al.}(2019){Xue}, {Remijan}, {Brogan}, {Hunter}, {Herbst},
  \& {McGuire}}]{Xue19}
{Xue}, C., {Remijan}, A.~J., {Brogan}, C.~L., {et~al.} 2019, \apj, 882, 118

\end{thebibliography}



\end{document}